\documentclass[preprint,3p,times,final]{elsarticle}
\usepackage{graphicx,amssymb}
\usepackage[fleqn]{amsmath}
\journal{Physica A}
\usepackage{color}

\biboptions{sort&compress}

\begin{document}

\begin{frontmatter}

\title{Exactly solved mixed spin-(1,1/2) Ising-Heisenberg distorted diamond chain}
\author[UPJS,ICMP]{Bohdan Lisnyi\fnref{ack1}}
\ead{lisnyj@icmp.lviv.ua}
\fntext[ack1]{B.L. acknowledges the financial support provided by the National Scholarship Programme of the Slovak Republic for the Support of Mobility of
Students, PhD Students, University Teachers, Researchers and Artists.}
\author[UPJS]{Jozef Stre\v{c}ka\corref{cor1}\fnref{ack2}}
\ead{jozef.strecka@upjs.sk}
\fntext[ack2]{J.S. acknowledges the financial support provided by the grant of The Ministry of Education, Science, Research and Sport of the Slovak Republic
under the contract Nos. VEGA 1/0331/15 and VEGA 1/0043/16, as well as, by grants of the Slovak Research and Development Agency provided under Contract Nos. APVV-0097-12 and APVV-14-0073.}
\address[UPJS]{Institute of Physics, Faculty of Science, P. J. \v{S}af\'{a}rik University, Park Angelinum 9, 040 01 Ko\v{s}ice, Slovak Republic}
\address[ICMP]{Institute for Condensed Matter Physics, National Academy of Sciences of Ukraine, 1 Svientsitskii Street, 79011 Lviv, Ukraine}
\cortext[cor1]{Corresponding author.}

\begin{abstract}
The mixed spin-(1,1/2) Ising-Heisenberg model on a distorted diamond chain with the spin-1 nodal atoms and the spin-1/2 interstitial atoms is exactly solved by the transfer-matrix method. An influence of the geometric spin frustration and the parallelogram distortion on the ground state, magnetization, susceptibility and specific heat of the mixed-spin Ising-Heisenberg distorted diamond chain are investigated in detail. It is demonstrated that the zero-temperature magnetization curve may involve intermediate plateaus just at zero and one-half of the saturation magnetization. The temperature dependence of the specific heat may have up to three distinct peaks at zero magnetic field and up to four distinct peaks at a non-zero magnetic field. The origin of multipeak thermal behavior of the specific heat is comprehensively studied.
\end{abstract}


\end{frontmatter}

\section{Introduction}

During the last few years, a considerable research interest has been devoted to the frustrated magnetism of
diamond spin chains, which was initiated by several unusual magnetic features of the natural mineral
azurite Cu$_3$(CO$_3$)$_2$(OH)$_2$ such as for instance an existence of the one-third magnetization plateau
in a low-temperature magnetization curve \cite{kik04,kik05,kik06,rul08,jes11,hon11}. However, it turns out that one
has to employ sophisticated first-principle density-functional calculations in combination with the extensive
state-of-the-art numerical calculations in order to provide a comprehensive description of the overall magnetic
behavior of the azurite \cite{jes11,hon11}. Compared to this, some exactly solved Ising-Heisenberg spin chains
afford a plausible quantitative description of the magnetic behavior of real spin-chain materials after modest calculations
in spite of a certain over-simplification of the physical reality  \cite{s0,s3,exp10,sah12,str12,han13,oha13,pssb14}.

In this regard, a lot of attention has been paid to a rigorous treatment of various versions of the spin-1/2 Ising-Heisenberg diamond chain \cite{can06,can09,lis3,ana12,roj12,ana13,gal13,gal14,anan14,faizi14,faizi15,ohanyan15}, the spin-1 Ising-Heisenberg diamond chain \cite{vahan16,ssc14,ssc15-01,jpcm16} and also the correlated spin-electron diamond chain \cite{dos3,dos3a,lis11-1,lis13-1,nalbandyan,torrico}.
On the other hand, much less attention has been devoted to exactly solvable cases of the mixed-spin Ising-Heisenberg diamond chains \cite{can09}.
Among this extensive class of geometrically frustrated spin chains the mixed spin-(1/2,1) Ising-Heisenberg diamond chain with the spin-1/2 nodal atoms
and the spin-1 interstitial (decorating) atoms was the most intensively studied \cite{can06,can09,roj11,jas04,ssc15-02}.
Recently, Zihua Xin \textit{et al.} \cite{xin12} have explored the mixed-spin Ising diamond chain with the spin-1 nodal atoms and the spin-1/2 interstitial
atoms by employing Monte Carlo simulations. It has been argued in Ref. \cite{xin12} that the low-temperature magnetization process of the mixed spin-(1,1/2)
Ising diamond chain displays peculiar magnetization plateaus, which have been later refuted by the exact calculations based on the generalized decoration-iteration
transformation \cite{jmmm13}. Besides, we have recently furnished a rigorous proof that the striking magnetization plateaus proposed in Ref. \cite{xin12}
cannot appear neither in the generalized mixed spin-(1,1/2) Ising-Heisenberg diamond chain, which accounts for the single-ion anisotropy
acting on the nodal spin-1 atoms \cite{jmmm15}.

The main purpose of this work is to examine the ground state and basic thermodynamic properties of the mixed spin-(1,1/2) Ising-Heisenberg model on a distorted diamond chain,
which accounts for asymmetry of the Ising couplings along the sides of elementary diamond plaquette distorted to parallelogram. The distorted version of the mixed spin-(1,1/2) Ising-Heisenberg diamond chain reduces to the usual mixed spin-(1,1/2) Ising-Heisenberg diamond chain when the asymmetry of the Ising couplings along the diamond sides vanishes, while the
mixed spin-(1,1/2) Ising-Heisenberg doubly decorated chain is recovered under the extreme case of the asymmetry. The ground state, magnetization process and basic thermodynamic characteristics
(entropy, specific heat, susceptibility) of the mixed spin-(1,1/2) Ising-Heisenberg distorted diamond chain will be exactly calculated within the framework of the transfer-matrix method.
In particular, we will explore how a mutual competition between the geometric spin frustration and the coupling asymmetry will affect the overall magnetic behavior.

The organization of this paper is as follows. The model and basic steps of the exact method will be clarified in Sec. \ref{model}. The most interesting results for the ground-state phase diagrams and thermodynamic properties will be discussed in Sec. \ref{result}. Finally, the paper ends up with several concluding remarks and future outlooks mentioned in Sec. \ref{conclusion}.

\section{Model and its exact solution}
\label{model}

\begin{figure}
\begin{center}
\includegraphics[width=0.5\columnwidth]{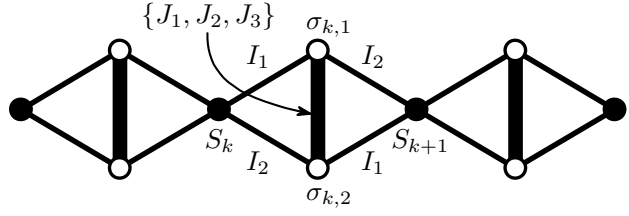}
\end{center}
\vspace{-0.5cm} \caption{A fragment from the mixed-spin Ising-Heisenberg distorted diamond chain.
The nodal ($S_{k}$, $S_{k+1}$) Ising spins and interstitial ($\sigma_{k,1}$, $\sigma_{k,2}$) Heisenberg spins
belonging to the $k$th primitive unit cell are marked. }
\label{fig1}
\end{figure}

Let us begin by considering the mixed spin-(1,1/2) Ising-Heisenberg distorted diamond chain in
a presence of the external magnetic field.
The magnetic structure of the investigated model system is schematically illustrated in Fig.~\ref{fig1}
together with its primitive unit cell.
As one can see, the primitive unit cell in a shape of diamond spin cluster involves two nodal Ising spins
$S_{k}$ and $S_{k+1}$ along with two interstitial Heisenberg spins $\hat\sigma_{k,1}$ and $\hat\sigma_{k,2}$.
The total Hamiltonian for the mixed spin-(1,1/2) Ising-Heisenberg distorted diamond chain in a presence of the external
magnetic field reads
\begin{eqnarray}
\hat {\cal H} &=& \sum_{k=1}^{N}
S_{k} \left(I_1 \hat\sigma^z_{k,1} + I_2\hat\sigma^z_{k,2} + I_2\hat\sigma^z_{k-1,1} + I_1 \hat\sigma^z_{k-1,2} \right)
+ \sum_{k=1}^{N} \left( J_1 \hat\sigma^x_{k,1} \hat\sigma^x_{k,2} + J_2 \hat\sigma^y_{k,1} \hat\sigma^y_{k,2}
+ J_3 \hat\sigma^z_{k,1} \hat\sigma^z_{k,2} \right)
\nonumber\\
&& {} - \sum_{k=1}^{N} h \left( S_{k} + \hat\sigma^z_{k,1} + \hat\sigma^z_{k,2} \right),
\label{htot}
\end{eqnarray}
which involves the nodal Ising spins $S_k = \pm 1, 0$ and the interstitial Heisenberg spins $\sigma_{k,i} = 1/2$ ($i = 1,2$)
by assuming the periodic boundary conditions $\hat\sigma^{\alpha}_{0,1} \equiv \hat\sigma^{\alpha}_{N,1}$,
$\hat\sigma^{\alpha}_{0,2} \equiv \hat\sigma^{\alpha}_{N,2}$ ($\alpha = x,y,z$).
The interaction constants $I_1$ and $I_2$ label the nearest-neighbor interactions between the nodal Ising spins and
interstitial Heisenberg spins along sides of the primitive diamond unit cell,
while the coupling constants $J_1$, $J_2$ and $J_3$ determine the spatially anisotropic XYZ interaction between
the nearest-neighbor interstitial Heisenberg spins from the same primitive cell.
Finally, the Zeeman's term $h$ determines the magnetostatic energy of the nodal Ising spins and
interstitial Heisenberg spins in the external magnetic field.
It should be mentioned that the particular case $I_2=0$ (or $I_1=0$) of the Hamiltonian (\ref{htot})
corresponds to the mixed spin-(1,1/2) Ising-Heisenberg doubly decorated chain.

For  further manipulations, it is quite advisable to rewrite the total Hamiltonian (\ref{htot}) as a sum over
cell Hamiltonians
\begin{equation}
\hat {\cal H} = \sum_{k=1}^N \hat {\cal H}_k,
\label{hk}
\end{equation}
whereas the cell Hamiltonian $\hat {\cal H}_k$ involves all the interaction terms of the $k$th diamond unit cell
\begin{eqnarray}
\hat {\cal H}_k &=& J_1 \hat\sigma^x_{k,1} \hat\sigma^x_{k,2} + J_2 \hat\sigma^y_{k,1} \hat\sigma^y_{k,2}
+ J_3 \hat\sigma^z_{k,1} \hat\sigma^z_{k,2}
+ I_1 \left( S_{k}\hat\sigma^z_{k,1} + \hat\sigma^z_{k,2}S_{k+1} \right)
+ I_2 \left( S_{k}\hat\sigma^z_{k,2} + \hat\sigma^z_{k,1}S_{k+1} \right)
\nonumber \\
&& {} - h \left(\hat\sigma^z_{k,1} + \hat\sigma^z_{k,2}\right) - \frac{h}{2} \left(S_{k} + S_{k+1}\right).
\label{cell}
\end{eqnarray}
Due to the fact that the cell Hamiltonians $\hat {\cal H}_k$ commute between themselves,
$\left[\hat {\cal H}_k, \hat {\cal H}_n \right]=0$,
the partition function of the mixed spin-(1,1/2) Ising-Heisenberg diamond chain can be written in this form
\begin{eqnarray}
{\cal Z} \equiv \mbox{Tr} \exp \left(-\beta \hat {\cal H} \right)
= \sum_{\{S_k \}} \prod_{k=1}^N {\cal Z}_k(S_{k},S_{k+1}),
\label{pfd}
\end{eqnarray}
where the symbol $\sum_{\{S_k \}}$ marks a summation over all possible spin configurations of the nodal Ising spins
and
\begin{eqnarray}
{\cal Z}_k(S_{k},S_{k+1}) &=& \mbox{Tr}_{\{\sigma_{k,1}, ~\sigma_{k,2}\}} \exp \left(-\beta \hat {\cal H}_k \right)
\label{BF}
\end{eqnarray}
is the effective Boltzmann's factor obtained after tracing out spin degrees of freedom of two Heisenberg spins from the $k$-th primitive cell
[$\beta=1/(k_{\rm B} T)$, $k_{\rm B}$ is the Boltzmann's constant, $T$ is the absolute temperature].
To proceed further with a calculation, one necessarily needs to evaluate the effective Boltzmann's factor ${\cal Z}_k(S_{k},S_{k+1})$ given by Eq. (\ref{BF}).
For this purpose, it is quite advisable to pass to the matrix representation of the
cell Hamiltonian $\hat {\cal H}_k$ in the basis spanned over four available states of two Heisenberg spins
$\sigma^z_{k,1}$ and $\sigma^z_{k,2}$:
\begin{eqnarray}
|\!\uparrow, \uparrow \rangle_{k} = \left |\uparrow \right \rangle_{k,1} \left |\uparrow \right \rangle_{k,2},
\quad
|\! \downarrow, \downarrow \rangle_{k} = \left |\downarrow \right \rangle_{k,1} \left |\downarrow \right \rangle_{k,2},
\quad
|\! \uparrow, \downarrow \rangle_{k} = \left |\uparrow \right \rangle_{k,1} \left |\downarrow \right \rangle_{k,2},
\quad
|\!\downarrow, \uparrow \rangle_{k} = \left |\downarrow \right \rangle_{k,1} \left |\uparrow \right \rangle_{k,2},
\label{B}
\end{eqnarray}
whereas $|\!\!\uparrow\rangle_{k,i}$ and $|\!\!\downarrow\rangle_{k,i}$ denote two eigenvectors of the spin operator
$\hat\sigma^z_{k,i}$ with the respective eigenvalues $\sigma^z_{k,i} = 1/2$ and $-1/2$.
After a straightforward diagonalization of the cell Hamiltonian $\hat {\cal H}_k$ one obtains the following four
eigenvalues:
\begin{eqnarray}
{\cal E}_{k\,1,2}  &=& - \frac{h}{2} \left(S_{k} + S_{k+1} \right) + \frac{J_3}{4}
\pm \sqrt{\left(\frac{J_1 - J_2}{4}\right)^2  + \left [\frac{I_1 + I_2}{2} \left(S_{k} + S_{k+1}\right) - h\right ]^2} ~,
\nonumber \\
{\cal E}_{k\,3,4} &=& - \frac{h}{2} \left(S_{k} + S_{k+1} \right) - \frac{J_3}{4}
\pm \sqrt{\left(\frac{J_1 + J_2}{4}\right)^2  + \left (\frac{I_1 - I_2}{2}\right)^2 \left(S_{k} - S_{k+1}\right)^2} ~.
\label{Ek}
\end{eqnarray}
Now, one may simply use the eigenvalues (\ref{Ek}) in order to calculate the Boltzmann's factor (\ref{BF})
according to the relation
\begin{eqnarray}
{\cal Z}_k(S_{k},S_{k+1})&=&
2\exp \left[\frac{\beta h}{2} \left(S_{k} + S_{k+1} \right)\right]
\left[ \exp \left(\frac{\beta J_3}{4} \right)
\cosh \left(\beta \sqrt{\left(\frac{J_1 + J_2}{4}\right)^2  +
\left(\frac{I_1 - I_2}{2}\right)^2 \left(S_{k} - S_{k+1}\right)^2} \right) \right.
\nonumber \\
&& \left. {} + \exp \left(-\frac{\beta J_3}{4}\right)
\cosh \left(\beta \sqrt{\left(\frac{J_1 - J_2}{4}\right)^2  +
\left[\frac{I_1 + I_2}{2} \left(S_{k} + S_{k+1}\right) - h\right]^2} \right)
\right].
\label{bf1}
\end{eqnarray}

The Boltzmann's factor (\ref{bf1}) can be subsequently replaced through the generalized decoration-iteration
transformation \cite{fis59,roj09,str10,bell13} similarly as we have done this before \cite{jmmm13,jmmm15}.
But since this Boltzmann's factor is essentially a transfer matrix (see eq. (\ref{pfd})),
we can directly apply the transfer-matrix method \cite{bax82,aps15}.
For convenience we define the elements $V_{ij}$ of the transfer matrix ${\mathbf V}$  as follows:
\begin{eqnarray}
V_{ij} \equiv {\cal Z}_k \left( S_{k}{=}i{-}2,~S_{k+1}{=}j{-}2 \right) &=&
2\exp \left[\frac{\beta h}{2} \left(i + j - 4 \right)\right]
\left[ \exp \left(\frac{\beta J_3}{4} \right)
\cosh\left(\beta\sqrt{\left(\frac{J_1 + J_2}{4}\right)^2  + \left(\frac{I_1 - I_2}{2}\right)^2 \left(i - j\right)^2}\right)
\right.
\nonumber \\
&& \left. {} + \exp \left(-\frac{\beta J_3}{4}\right)
\cosh\left(\beta\sqrt{\left(\frac{J_1 - J_2}{4}\right)^2  + \left[\frac{I_1 + I_2}{2} \left(i + j - 4 \right) - h\right]^2} \right)
\right].
\label{trm}
\end{eqnarray}
The calculation of the partition function thus requires finding of the eigenvalues of the transfer matrix
\begin{eqnarray}
{\mathbf V} = \left( \begin{array}{lll}
V_{11} & ~V_{12} & ~V_{13} \\
V_{21} & ~V_{22} & ~V_{23} \\
V_{31} & ~V_{32} & ~V_{33}
\end{array}
\right).
\nonumber
\end{eqnarray}
The eigenvalues of this matrix are given by the roots of cubic characteristic equation
\begin{eqnarray}
\lambda^3 + K_2\lambda^2 + K_1\lambda + K_0 = 0,
\label{CE}
\end{eqnarray}
where
\begin{eqnarray}
K_2 &=& - \frac{1}{3} \left( V_{11} + V_{22} + V_{33}\right), \qquad
K_1 = V_{11} V_{22} + V_{11} V_{33}  + V_{22} V_{33} - V_{12}^2 - V_{13}^2 - V_{23}^2,
\nonumber\\
K_0 &=&  V_{11}V_{23}^2 + V_{22}V_{13}^2 + V_{33}V_{12}^2 - V_{11}V_{22}V_{33} - 2V_{12}V_{13}V_{23}.
\nonumber
\end{eqnarray}
The expressions for the three transfer-matrix eigenvalues are as follows (see, e.g., \cite{korn})
\begin{eqnarray}
\lambda_i = |K_2| + 2 \sqrt{p} \cos \left[\phi + \frac{2\pi}{3} (i-1) \right], \quad i=1,2,3,
\label{RCE}
\end{eqnarray}
where
\begin{eqnarray}
p &=& \frac{1}{3} \left( V_{12}^2 + V_{13}^2 + V_{23}^2 \right) +
\frac{1}{18} \left[\left(V_{11} - V_{22}\right)^2 + \left(V_{11} - V_{33}\right)^2 + \left(V_{22} - V_{33}\right)^2 \right],
\nonumber \\
\phi &=& \frac{1}{3} \arccos \left(\frac{-q}{\sqrt{p^3}}\right), \qquad\quad
q = \frac{1}{2} \left( |K_2|^3 - 3|K_2|p + K_0\right).
\nonumber
\end{eqnarray}
As a result, the partition function ${\cal Z}$ given by Eq. (\ref{pfd}) is determined by the transfer-matrix eigenvalues $\lambda_1$, $\lambda_2$, $\lambda_3$ through the formula:
\begin{eqnarray}
{\cal Z} = \mbox{Tr} \, {\mathbf V}^N  = \lambda_1^N + \lambda_2^N + \lambda_3^N.
\label{pff}
\end{eqnarray}

Exact results for other thermodynamic quantities follow quite straightforwardly from the formula (\ref{pff}) for the partition function ${\cal Z}$.
In the thermodynamic limit $N \to \infty$, the Gibbs free energy per unit cell can be evaluated from the formula
\begin{eqnarray}
g = -\frac{1}{\beta} \lim_{N \to \infty} \frac{1}{N} \ln {\cal Z} = -\frac{1}{\beta} \ln \lambda_{0},
\label{frebege}
\end{eqnarray}
where $\lambda_{0} = {\rm max} \{ \lambda_1, \lambda_2, \lambda_3 \}$ is the largest transfer-matrix eigenvalue.
It is easy to see from Eq. (\ref{trm}) that $V_{ij}>0$ for all non-zero temperatures ($T>0$).
This means that the transfer matrix ${\mathbf V}$ is positive matrix for all non-zero temperatures.
The Perron-Frobenius theorem (see, e.g., \cite{meyer}) implies that the transfer matrix has a positive largest eigenvalue,
which is a simple root of the characteristic equation (\ref{CE}), and any other eigenvalue is strictly smaller that it in absolute value.
The largest eigenvalue among the three transfer-matrix eigenvalues (\ref{RCE}) is then given by
\begin{eqnarray}
\lambda_0 = |K_2| + 2 \sqrt{p} \cos \left(\phi \right).
\label{L0}
\end{eqnarray}

The entropy $s$ and the specific heat $c$ per unit cell can be subsequently calculated from the formulas
\begin{eqnarray}
s = k_{\rm{B}} \beta^2  \frac{\partial g}{\partial \beta}
= k_{\rm{B}} \left( \ln \lambda_{0}
- \frac{\beta}{\lambda_{0}} \frac{\partial \lambda_{0}}{\partial \beta}\right) , \qquad
c = - \beta  \frac{\partial s}{\partial \beta}
= k_{\rm{B}} \beta^2   \left[\frac{1}{\lambda_{0}} \frac{\partial^2 \lambda_{0}}{\partial \beta^2} -
\frac{1}{\lambda_{0}^2} \left(\frac{\partial \lambda_{0}}{\partial \beta}\right)^2 \right],
\nonumber
\end{eqnarray}
whereas the total magnetization $m$ and magnetic susceptibility $\chi$ readily follow from the relations
\begin{eqnarray}
m = - \frac{\partial g}{\partial h} = \frac{1}{\beta \lambda_{0}} \frac{\partial \lambda_{0}}{\partial h}, \qquad
\chi = \frac{\partial m}{\partial h} =
\frac{1}{\beta} \left[\frac{1}{\lambda_{0}} \frac{\partial^2 \lambda_{0}}{\partial h^2} -
\frac{1}{\lambda_{0}^2} \left(\frac{\partial \lambda_{0}}{\partial h}\right)^2 \right].
\nonumber
\end{eqnarray}
It is quite convenient to get the derivatives $\frac{\partial \lambda_{0}}{\partial \beta}$, $\frac{\partial^2 \lambda_{0}}{\partial \beta^2}$,
$\frac{\partial \lambda_{0}}{\partial h}$, $\frac{\partial^2 \lambda_{0}}{\partial h^2}$ from Eq. (\ref{CE}) rather than from the final formula (\ref{L0}).

\section{Results and discussions}
\label{result}

Now, let us proceed to a discussion of the most interesting results obtained for the mixed spin-(1,1/2) Ising-Heisenberg distorted diamond chain with the antiferromagnetic Ising interactions $I_1>0$ and $I_2>0$. To reduce number of free parameters, we will further assume the XXZ Heisenberg interaction  $J_{1} = J_{2} = J \Delta$, $J_{3} = J$, which may be either antiferromagnetic or ferromagnetic in character ($\Delta$ determines a spatial anisotropy in this interaction). Without loss of generality, we may also assume that one of two considered Ising couplings is stronger than the other one $I_1 \geq I_2$ and to introduce the difference between both Ising coupling constants $\delta I = I_1 - I_2>0$. Subsequently, one gets the following four-dimensional parameter space spanned over the dimensionless interaction parameters
\[
\tilde{J}=\frac{J}{I_1},  \qquad \Delta, \qquad \tilde{h}=\frac{h}{I_1}, \qquad \delta \tilde{I}=\frac{\delta I}{I_1}=1-\frac{I_2}{I_1}.
\]
The parameters $\tilde{J}$ and $\Delta$ determine a relative strength of the XXZ Heisenberg interaction, while the parameter $\tilde{h}$ stands for a relative strength of the external magnetic field. Last, the parameter $\delta \tilde{I}$ restricted to the interval $\delta \tilde{I} \in \textcolor{red}{[}0,1\textcolor{red}{]}$ has a physical sense of the distortion parameter, because it determines a relative difference between two Ising coupling constants $I_1$ and $I_2$ due to the parallelogram distortion.

\subsection{Ground state for antiferromagnetic Heisenberg interaction}

The ground state of the mixed spin-(1,1/2) Ising-Heisenberg diamond chain can be trivially connected to
the lowest-energy eigenstate of the cell Hamiltonian (\ref{cell}) obtained by taking into account all nine states of the nodal Ising spins $S_{k}$ and $S_{k+1}$ entering into
the respective eigenvalues (\ref{Ek}). Depending on a mutual competition between the parameters $\Delta$, $\tilde{J}>0$, $\delta \tilde{I}$ and $\tilde{h}$
one finds in total four different ground states: the saturated paramagnetic (SPA) state, the monomer-dimer (MD) state, the classical antiferromagnetic (AF) state, and the quantum antiferromagnetic (QAF) state given by the eigenvectors
\begin{eqnarray}
|\mbox{SPA} \rangle &=& \prod\limits_{k=1}^N| + \rangle_k \otimes |\!\uparrow, \uparrow \rangle_{k},
\qquad \qquad
|\mbox{MD} \rangle = \prod\limits_{k=1}^N | + \rangle_k \otimes
\frac{1}{\sqrt{2}} \Big(|\!\uparrow, \downarrow \rangle_{k} - |\!\downarrow, \uparrow \rangle_{k} \Big),
\nonumber\\
|\mbox{AF} \rangle  &=& \left \{\begin{array}{l}
\prod\limits_{k=1}^N | - \rangle_k \otimes |\!\uparrow, \uparrow \rangle_{k}
\\
\prod\limits_{k=1}^N | + \rangle_k \otimes |\!\downarrow, \downarrow \rangle_{k}
\end{array}\right. ,
\quad \qquad
|\mbox{QAF} \rangle = \left\{\begin{array}{l}
\prod\limits_{k=1}^N \left|[-]^k \right\rangle_k \otimes
\left(A_{[-]^{k+1}} |\!\uparrow, \downarrow \rangle_{k} - A_{[-]^k} |\!\downarrow, \uparrow \rangle_{k} \right)
\\
\prod\limits_{k=1}^N \left|[-]^{k+1} \right\rangle_k \otimes
\left(A_{[-]^k} |\!\uparrow, \downarrow \rangle_{k} - A_{[-]^{k+1}} |\!\downarrow, \uparrow \rangle_{k} \right)
\end{array} \right. .
\label{GS}
\end{eqnarray}
In above, the ket vector $|\pm\rangle_k$ determines the state of the nodal Ising spin $S_k = \pm 1$,
the symbol $[-]^k \in \{-,+\}$ marks the sign of the number $(-1)^k$,
the spin states relevant to two Heisenberg spins from the $k$th primitive cell are determined by the notation (\ref{B}),
and the probability amplitudes $A_{\pm}$ are explicitly given by the expressions
\begin{equation}
A_{\pm}=\frac{1}{\sqrt{2}}\sqrt{1 \pm \frac{2\delta \tilde{I}}
{\sqrt{\left(\tilde{J} \Delta\right)^2 + 4\delta \tilde{I}^2}}}.
\label{PA}
\end{equation}
The eigenenergies per primitive cell that correspond to the respective ground states (\ref{GS}) are given as follows
\begin{eqnarray}
\tilde{\cal E}_{\rm{SPA}} = \frac{\tilde{J}}{4} + 2 - \delta\tilde{I} - 2\tilde{h},
\quad
\tilde{\cal E}_{\rm{MD}} = -\frac{\tilde{J}}{4} - \frac{1}{2}\tilde{J}\Delta  - \tilde{h},
\quad
\tilde{\cal E}_{\rm{AF}} = \frac{\tilde{J}}{4} - 2 + \delta\tilde{I},
\quad
\tilde{\cal E}_{\rm{QAF}} = -\frac{\tilde{J}}{4} -
\frac{1}{2}\sqrt{\left(\tilde{J} \Delta \right)^2 + 4\delta \tilde{I}^2}.
\label{EE}
\end{eqnarray}
Let us make a few comments on two notable spin arrangements emerging within the QAF and MD ground states. Apparently, one encounters within the QAF ground state a singlet-like state of the Heisenberg spin pairs, which is characterized by a quantum superposition of two antiferromagnetic states $|\! \uparrow, \downarrow \rangle_{k}$ and $|\! \downarrow,  \uparrow \rangle_{k}$ occurring according to Eq. (\ref{PA}) with two different occurrence probabilities. As a result, there appears an effective staggered magnetic moment on the Heisenberg spin pairs, which is subsequently transferred also to the nodal Ising spins provided that there is some asymmetry in both Ising couplings $\delta\tilde{I} \neq 0$ due to the parallelogram distortion. Contrary to this, the perfect singlet-dimer state of the Heisenberg spin pairs within the MD ground state is responsible for a geometric spin frustration of the nodal Ising spins, which are consequently forced by the external magnetic field to align towards the magnetic-field direction. In this regard, the MD ground state should manifest itself in a respective magnetization curve as an intermediate plateau at one-half of the saturation magnetization.

\begin{figure}
\begin{center}
\includegraphics[width=0.95\textwidth]{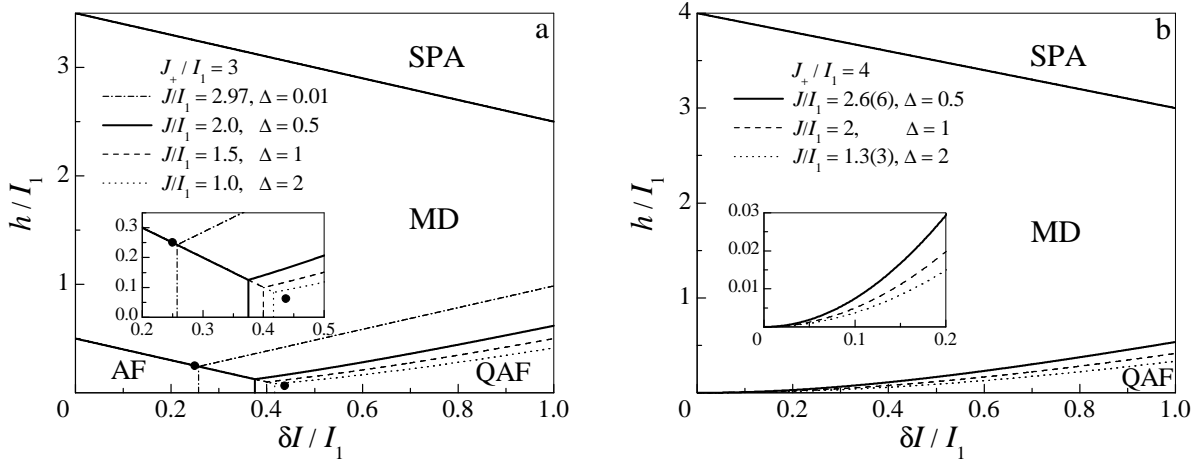}
\end{center}
\vspace{-0.5cm}
\caption{The ground-state phase diagram in the $\delta\tilde{I} - \tilde{h}$ plane for a few typical values of the antiferromagnetic Heisenberg coupling constant ($\tilde{J}>0$, $\Delta$), which are consistent with two different values of the parameter $\tilde{J}_{+}$: (a) $\tilde{J}_{+} = 3$; (b) $\tilde{J}_{+} = 4$. The bullet symbols $\bullet$ shown in Fig. \ref{fig2}a allocate two limiting positions of the triple point,
at which the ground states AF, QAF, and MD coexist together on assumption that $\tilde{J}_{+} = 3$.}
\label{fig2}
\end{figure}
The ground-state phase diagram in the $\delta \tilde{I} - \tilde{h}$ plane can have topology of two different types depending on a strength of the parameter $\tilde{J}_{+}=J_{+}/I_1=\tilde{J}(1+\Delta)$ connected to the Heisenberg interaction (see Fig.~\ref{fig2}). The first type of the ground-state phase diagram involving all four available ground states SPA, MD, AF and QAF is found on assumption that $\tilde{J}_{+}<4$ (see Fig.~\ref{fig2}a). Under this condition, the AF and QAF ground states coexist together at the following values of the distortion parameter
\begin{eqnarray}
\delta\tilde{I} = \delta\tilde{I}_{\textrm{AF}\,|\,\textrm{QAF}} \equiv
\frac{4 - \tilde{J}_{+}}{4}\left (1 + \frac{\tilde{J}_{+} \Delta}{4(1 + \Delta) - \tilde{J}_{+}}\right).
\end{eqnarray}
If the Heisenberg coupling constant $\tilde{J}_{+}$ is fixed, then, the distortion parameter is restricted to the interval $\left(1-\tilde{J}_{+}/4, 1- \left(\tilde{J}_{+}/4\right)^2 \right)$ at the coexistence line $\delta\tilde{I}=\delta\tilde{I}_{\textrm{AF}\,|\,\textrm{QAF}}$ as the exchange anisotropy $\Delta$ varies from 0 to $\Delta \rightarrow \infty$. The ground-state phase diagram depicted in Fig.~\ref{fig2}a additionally contains a special triple point given by the coordinates
$\left[\delta\tilde{I}_{\textrm{AF}\,|\,\textrm{QAF}}, 2 - \delta\tilde{I}_{\textrm{AF}\,|\,\textrm{QAF}} - \tilde{J}_{+}/2\right]$,
at which the three ground states AF, QAF and MD coexist together. For the fixed value of the interaction parameter $\tilde{J}_{+}$ the position of
the triple point moves with the increase of parameter $\Delta$ along the line from the point $\left[1-\tilde{J}_{+}/4,1-\tilde{J}_{+}/4\right]$ at $\Delta=0$ up to the point $\left[1-(\tilde{J}_{+}/4)^2,(1-\tilde{J}_{+}/4)^2 \right]$ reached in the $\Delta \rightarrow \infty$ limit (see bullet symbols in Fig.~\ref{fig2}a).

The second type of the ground-state phase diagram can be detected for $\tilde{J}_{+} \geq 4$ and it includes just three different ground states SPA, MD and QAF (see Fig.~\ref{fig2}b). In this particular case the phase boundary between the QAF and MD  ground states starts at the point with the coordinates $[0,0]$, because the AF state is absent in the ground-state phase diagram. The special point $[0,0]$ of the ground-state phase diagram, which corresponds to the symmetric case at zero magnetic field, entails for $\tilde{J}_{+} > 4$ the highly frustrated (FRU) ground state
\[
|\mbox{FRU} \rangle = \prod\limits_{k=1}^N
\Big(\left|\pm\right\rangle_k \mbox{or} \left|0\right\rangle_k \Big) \otimes
\frac{1}{\sqrt{2}} \Big(|\!\uparrow, \downarrow \rangle_{k} - |\!\downarrow, \uparrow \rangle_{k} \Big),
\]
where the ket vector $|0\rangle_k$ determines the non-magnetic state of the nodal Ising spin $S_k = 0$. Hence, the FRU ground state has the residual entropy $s_{\rm{res}}= k_{\rm B}\ln 3$ reflecting the macroscopic degeneracy $3^N$, which stems from the degrees of freedom of the nodal Ising spins.
It should be also mentioned that the two ground states FRU and AF coexist together at the point $[0,0]$ for $\tilde{J}_{+} = 4$.

\begin{figure}
\begin{center}
\includegraphics[width=0.6\textwidth]{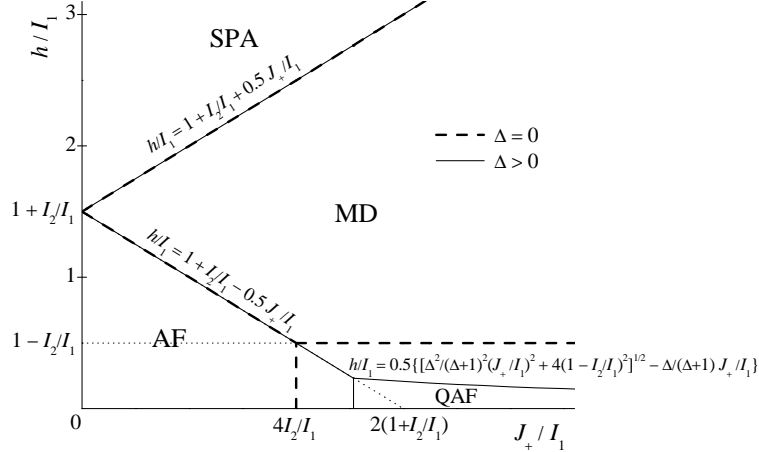}
\end{center}
\vspace{-0.7cm}
\caption{The ground-state phase diagram in the $\tilde{J}_{+} - \tilde{h}$ plane. The scale of the $\tilde{J}_{+}$-axis is expressed in terms of the interaction ratio $I_2/I_1$. Solid lines determine the ground-state phase boundaries for $\Delta>0$, along which the relevant mathematical formulas are also explicitly given. Broken lines illustrate the ground-state phase boundaries for the special case $\Delta=0$. The horizontal dotted line shows an imaginary continuation of the phase boundary to its intersection with the $\tilde{h}$-axis, while the oblique dotted line shows how a position of the triple coexistence point AF-QAF-MD moves along the $\tilde{J}_{+}$-axis.}
\label{fig3}
\end{figure}

The general form of the ground-state phase diagram can be obtained in the $\tilde{J}_{+}-\tilde{h}$ plane when the scale of the $\tilde{J}_{+}$-axis is expressed in terms of the interaction ratio $\tilde{I}_2=I_2/I_1$ between both Ising coupling constants (see Fig. \ref{fig3}). The phase boundary between the AF and QAF ground states then reads
\begin{eqnarray}
\tilde{J}_{+} = \tilde{J}_{+\,\textrm{AF}\,|\,\textrm{QAF}} \equiv
\left\{\begin{array}{ll}
\frac{8\tilde{I}_2}{1+\tilde{I}_2}, & ~\Delta=1
\\
\frac{2}{\Delta-1} \left( \sqrt{\left( 1-\tilde{I}_2 \right)^2 + 4 \tilde{I}_2 \Delta^2} -1 - \tilde{I}_2 \right), &
~\Delta \neq 1
\end{array} \right..
\nonumber
\end{eqnarray}
It is noteworthy that the FRU ground state does exist in the relevant ground-state phase diagram along the line given by $\tilde{h}=0$ and $\tilde{J}_{+}>4$ if the complete symmetry is recovered (i.e. $\tilde{I}_2=1$), because the QAF ground state then completely disappears from the ground-state phase diagram.

\subsection{Ground state for ferromagnetic Heisenberg interaction}

Depending on a mutual competition between the parameters $\Delta$, $\tilde{J}=-|\tilde{J}|<0$, $\delta \tilde{I}$ and $\tilde{h}$ one finds in total four different ground states for the particular case of the ferromagnetic Heisenberg interaction: the saturated paramagnetic (SPA) state, the monomer-dimer (MD1) state, the classical antiferromagnetic  (AF) state,
and the quantum antiferromagnetic (QAF1) state. Two new quantum ground states MD1 and QAF1 are given by the eigenvectors
\begin{eqnarray}
|\mbox{MD1} \rangle &=& \prod\limits_{k=1}^N | + \rangle_k \otimes
\frac{1}{\sqrt{2}} \Big(|\!\uparrow, \downarrow \rangle_{k} + |\!\downarrow, \uparrow \rangle_{k} \Big),
\qquad
|\mbox{QAF1} \rangle = \left\{\begin{array}{l}
\prod\limits_{k=1}^N \left|[-]^k \right\rangle_k \otimes
\left(A_{[-]^{k+1}} |\!\uparrow, \downarrow \rangle_{k} + A_{[-]^k} |\!\downarrow, \uparrow \rangle_{k} \right)
\\
\prod\limits_{k=1}^N \left|[-]^{k+1} \right\rangle_k \otimes
\left(A_{[-]^k} |\!\uparrow, \downarrow \rangle_{k} + A_{[-]^{k+1}} |\!\downarrow, \uparrow \rangle_{k} \right)
\end{array} \right.,
\label{GS1}
\end{eqnarray}
whereas they differ from the analogous ground states MD and QAF just by the symmetric (instead of antisymmetric) quantum superposition of the antiferromagnetic states $|\! \uparrow, \downarrow \rangle_{k}$ and $|\! \downarrow,  \uparrow \rangle_{k}$ of the Heisenberg spin pairs. The eigenenergies per primitive cell, which correspond to the respective ground states (\ref{GS1}), follow from
\begin{eqnarray}
\tilde{\cal E}_{\rm{MD1}} = -\frac{\tilde{J}}{4} - \frac{1}{2}|\tilde{J}|\Delta  - \tilde{h},
\qquad
\tilde{\cal E}_{\rm{QAF1}} = -\frac{\tilde{J}}{4} -
\frac{1}{2}\sqrt{\left(\tilde{J} \Delta \right)^2 + 4\delta \tilde{I}^2}.
\label{EE1}
\end{eqnarray}

\begin{figure}
\begin{center}
\includegraphics[width=0.9\textwidth]{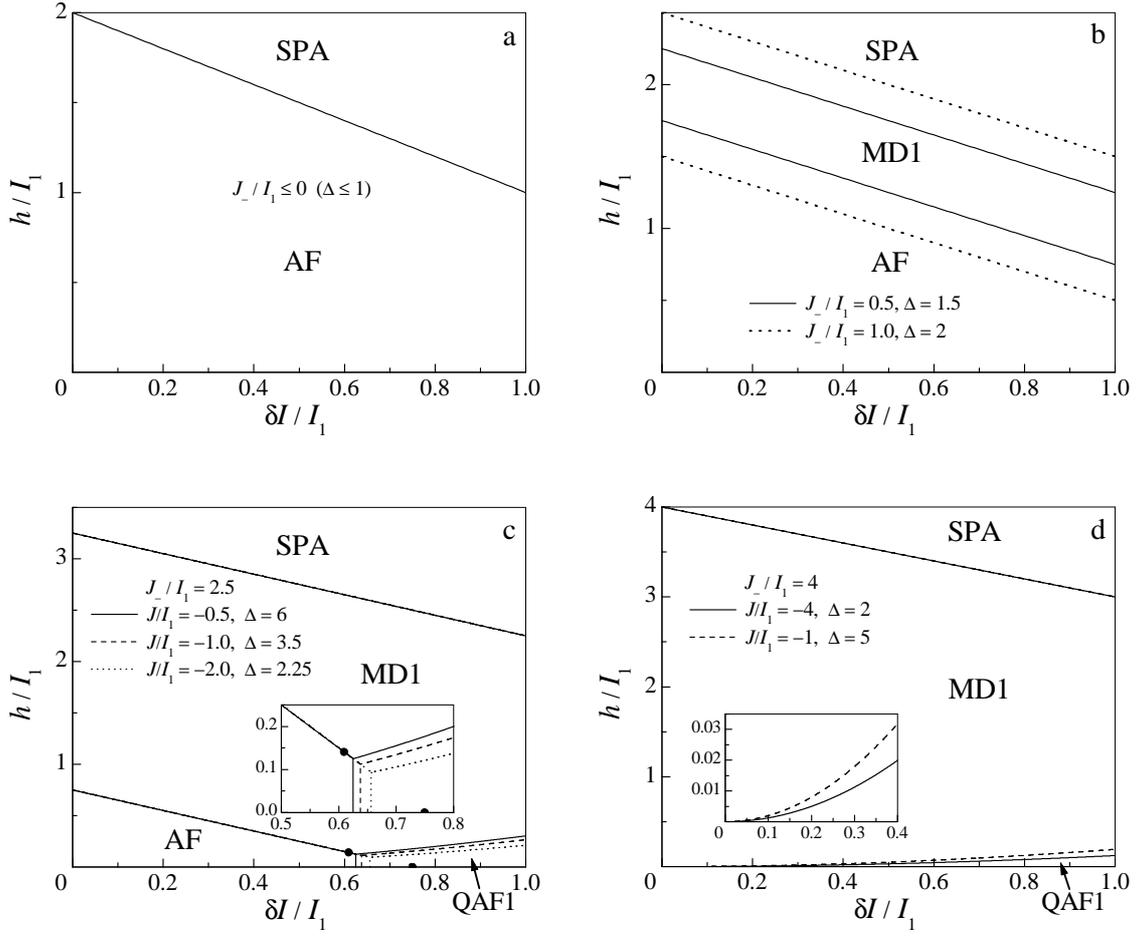}
\end{center}
\vspace{-0.5cm}
\caption{The ground-state phase diagram in the $\delta\tilde{I} - \tilde{h}$ plane for a few typical values of the ferromagnetic Heisenberg coupling constant ($\tilde{J}<0$, $\Delta$), which are consistent with four different topologies depending on a value of the parameter $\tilde{J}_{-}$: (a) $\tilde{J}_{-} \leq 0$; (b) $\tilde{J}_{-} = 0.5$ and $1.0$; (c) $\tilde{J}_{-} = 2.5$; (d) $\tilde{J}_{-} = 4.0$. The bullet symbols $\bullet$ shown in Fig. \ref{fig4}c allocate two limiting positions of the triple point, at which the ground states AF, QAF1, and MD1 coexist together on assumption that $\tilde{J}_{-} = 2.5$.}
\label{fig4}
\end{figure}
The ground-state phase diagram in the $\delta \tilde{I}-\tilde{h}$ plane can have four different topologies depending on a parameter $\tilde{J}_{-}= J_{-}/I_1= |\tilde{J}|(\Delta - 1)$ (see Fig.~\ref{fig4}). The first type of the ground-state phase diagram (Fig.~\ref{fig4}a) is found for
$\tilde{J}_{-} \leq 0$. It includes only two ground states SPA and AF. The second type of the ground-state phase diagram (Fig.~\ref{fig4}b) is realized for $0<\tilde{J}_{-} \leq 4/(\Delta + 1)$ and it involves three different ground states: SPA, AF, and MD1. The third type of the ground-state phase diagram (Fig.~\ref{fig4}c) involving all four available ground states SPA, MD1, AF, and QAF1 can be detected for $4/(\Delta + 1)<\tilde{J}_{-} < 4$.
The AF and QAF1 ground states are separated by the phase boundary
\begin{eqnarray}
\delta\tilde{I} = \delta\tilde{I}_{\textrm{AF}\,|\,\textrm{QAF1}} \equiv
\frac{4 - \tilde{J}_{-}}{4}\left (1 + \frac{\tilde{J}_{-}\Delta}{4 (\Delta - 1) + \tilde{J}_{-}} \right).
\end{eqnarray}
If the Heisenberg coupling constant $\tilde{J}_{-}$ is fixed, then, the distortion parameter is restricted to the interval $\left(1 - \left(\tilde{J}_{-}/4\right)^2, 2-\tilde{J}_{-}/2, \right)$ at the coexistence line $\delta\tilde{I}=\delta\tilde{I}_{\textrm{AF}\,|\,\textrm{QAF1}}$ as the exchange anisotropy varies from $\Delta \rightarrow \infty$ to $\Delta \rightarrow 1$. The ground-state phase diagram also contains a special triple point with the coordinates $\left[\delta\tilde{I}_{\textrm{AF}\,|\,\textrm{QAF1}},2 - \delta\tilde{I}_{\textrm{AF}\,|\,\textrm{QAF1}} - \tilde{J}_{-}/2\right]$, at which the AF, QAF1, MD1 phases coexist together (see the inset in Fig.~\ref{fig4}c).
For the fixed value of the interaction parameter $\tilde{J}_{-}$ the triple point moves
with the change of parameter $\Delta$ along the line from the point $\left[1-(\tilde{J}_{-}/4)^2,(1-\tilde{J}_{-}/4)^2 \right]$ at $\Delta \rightarrow \infty$ to the point $\left[2-\tilde{J}_{-}/2,0\right]$ at $\Delta \rightarrow 1$. The fourth type of the ground-state phase diagram (Fig.~\ref{fig4}d) emerges for $\tilde{J}_{-} \geq 4$ and it includes three different ground states: SPA, MD1, and QAF1. The phase boundary between the QAF1 and MD1 states starts from the special point $[0,0]$, which repeatedly corresponds to the highly frustrated (FRU1) ground state also for $\tilde{J}_{-} > 4$
\[
|\mbox{FRU1} \rangle = \prod\limits_{k=1}^N
\Big(\left|\pm\right\rangle_k \mbox{or} \left|0\right\rangle_k \Big) \otimes
\frac{1}{\sqrt{2}} \Big(|\uparrow, \downarrow \rangle_{k} + |\downarrow, \uparrow \rangle_{k} \Big).
\]
The two ground states FRU1 and AF coexist together at the point $[0,0]$ for $\tilde{J}_{-} = 4$. The FRU1 ground state has the residual entropy $s_{\rm{res}} = k_{\rm B}\ln 3$ reflecting the macroscopic degeneracy $3^N$, which comes from spin degrees of freedom of the nodal Ising spins. The only difference between two frustrated ground states FRU1 and FRU lies in symmetric vs. antisymmetric quantum superposition of two antiferromagnetic states $|\! \uparrow, \downarrow \rangle_{k}$ and $|\! \downarrow,  \uparrow \rangle_{k}$ of the Heisenberg spin pairs.

\begin{figure}[h!]
\begin{center}
\includegraphics[width=0.6\textwidth]{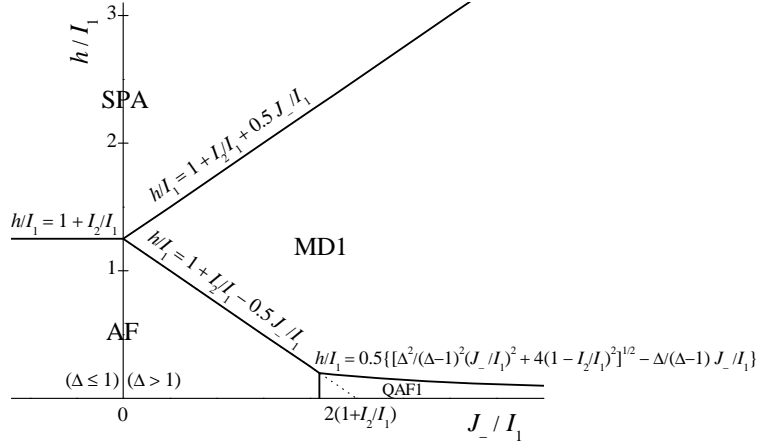}
\end{center}
\vspace{-0.7cm}
\caption{The ground-state phase diagram in the $\tilde{J}_{-} - \tilde{h}$ plane. The scale of the $\tilde{J}_{-}$-axis is expressed in terms of the interaction ratio $I_2/I_1$. Solid lines determine the ground-state phase boundaries, along which the relevant mathematical formulas are also explicitly given. The oblique dotted line shows how a position of the triple coexistence point AF-QAF1-MD1 moves along the $\tilde{J}_{-}$-axis.}
\label{fig5}
\end{figure}

The general form of the ground-state phase diagram can be obtained in the $\tilde{J}_{-}-\tilde{h}$ plane when the $\tilde{J}_{-}$-axis is expressed in terms of the interaction ratio $\tilde{I}_2 = I_2/I_1$ (Fig.~\ref{fig5}). Under this circumstance, the phase boundary between the AF and QAF1 ground states is given by
\begin{eqnarray}
\tilde{J}_{-} = \tilde{J}_{-\,\textrm{AF}\,|\,\textrm{QAF1}} \equiv
\frac{2}{\Delta+1} \left( \sqrt{\left( 1-\tilde{I}_2 \right)^2 + 4 \tilde{I}_2 \Delta^2} +1 + \tilde{I}_2 \right) .
\nonumber
\end{eqnarray}
It should be noticed that the FRU1 ground state does exist in the relevant ground-state phase diagram along the line given by $\tilde{h}=0$ and $\tilde{J}_{-}>4$ if the complete symmetry $\tilde{I}_2=1$ is recovered due to the absence of the QAF1 ground state.

\subsection{Thermodynamic properties}

Next, let us examine the magnetization process and other thermodynamic characteristics as a function of the temperature and the distortion parameter.
To illustrate all possible scenarios, we have selected the values of the Heisenberg coupling constants $\tilde{J}$ and $\Delta$ in order to fall into
the parameter region pertinent to the ground-state phase diagrams shown in Fig.~\ref{fig2}a and Fig.~\ref{fig4}c involving all available ground states.

\begin{figure}[htb]
\begin{center}
\includegraphics[width=0.9\textwidth]{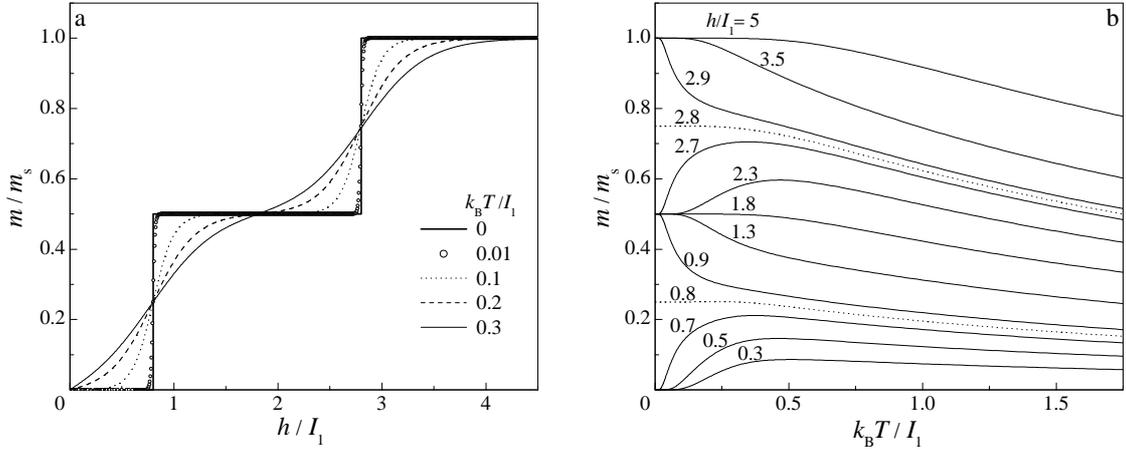}
\end{center}
\vspace{-0.5cm}
\caption{(a) The total magnetization normalized with respect to the saturation magnetization as a function of the magnetic field at a few different temperatures for
the particular case with the antiferromagnetic Heisenberg interaction $\tilde{J}=1$, $\Delta=1$ ($\tilde{J}_{+}=2$) and the distortion parameter $\delta\tilde{I}=0.2$.
(b) Thermal variations of the total magnetization for the same particular case with the antiferromagnetic Heisenberg interaction $\tilde{J}=1$, $\Delta=1$ and the distortion parameter $\delta\tilde{I}=0.2$ at several values of the magnetic field. The dotted lines correspond to the special values of the magnetic field $\tilde{h}=0.8$ and $2.8$, at which two different ground states coexist together.}
\label{fig6}
\end{figure}

The total magnetization is plotted in Fig.~\ref{fig6}a against the magnetic field  at a few different temperatures for the antiferromagnetic Heisenberg coupling $\tilde{J}=1$, $\Delta=1$ ($\tilde{J}_{+}=2$) and the distortion parameter $\delta\tilde{I}=0.2$, whereas the respective thermal dependences of the total magnetization at constant magnetic fields are displayed in Fig.~\ref{fig6}b. The zero-temperature dependence of the total magnetization depicted in Fig.~\ref{fig6}a exhibits two abrupt magnetization jumps closely connected with an existence
of two intermediate magnetization plateaus: the zero plateau corresponding to the AF ground state and the one-half plateau corresponding to the MD ground state. In accordance with this statement, temperature dependences of the total magnetization shown in Fig.~\ref{fig6}b asymptotically tend towards zero, one-half or unity as temperature goes to zero. The only two exceptions from this rule are the intermediate values of the total magnetization $m/m_s = 1/4$ and $3/4$, which can be achieved in the zero-temperature asymptotic limit on assumption that the magnetic field is fixed exactly
at the critical value corresponding to the field-induced transitions AF~$\leftrightarrow$~MD and MD~$\leftrightarrow$~SPA, respectively (see dotted lines in Fig.~\ref{fig6}b). Besides, it can be observed from Fig.~\ref{fig6}b that the total magnetization exhibits a relatively steep thermally-induced increase (decrease) at low enough temperatures when the magnetic field is selected slightly below (above) the critical field associated with the respective magnetization jump. The qualitatively same magnetization curve can be detected also for the other particular case $\delta\tilde{I}>\delta\tilde{I}_{\textrm{AF}\,|\,\textrm{QAF}}$, which drives the zero-field ground state towards the QAF phase instead of the AF phase. The only qualitative difference is that the total magnetization normalized with respect to the saturation magnetization asymptotically reaches in a zero-temperature limit the specific value $m/m_s = 1/(2 \sqrt{5}) \approx 0.2236$ when the critical field of the field-induced transition QAF~$\leftrightarrow$~MD is chosen.
\begin{figure}[htb]
\begin{center}
\includegraphics[width=0.95\columnwidth]{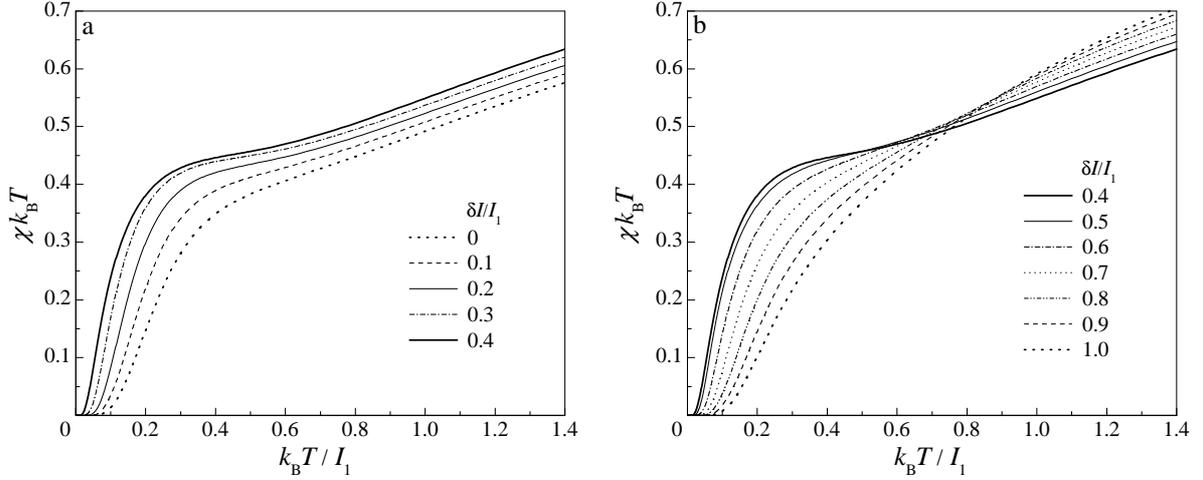}
\end{center}
\vspace{-0.5cm}
\caption{The zero-field susceptibility times temperature product as a function of temperature for the antiferromagnetic Heisenberg interaction $\tilde{J}=1.5$, $\Delta=1$ ($\tilde{J}_{+}=3$)
and several values of the distortion parameter: (a) $\delta \tilde{I}\leq\delta \tilde{I}_{\textrm{AF}\,|\,\textrm{QAF}}$; (b) $\delta \tilde{I}\geq\delta \tilde{I}_{\textrm{AF}\,|\,\textrm{QAF}}$.}
\label{fig7}
\end{figure}
The zero-field susceptibility times temperature product ($\chi T$) is presented in Fig.~\ref{fig7} as a function of the temperature for two particular cases:
when the distortion parameter $\delta \tilde{I}\leq\delta \tilde{I}_{\textrm{AF}\,|\,\textrm{QAF}}$ drives the investigated system towards the AF ground state
or when the distortion parameter  $\delta \tilde{I}\geq\delta \tilde{I}_{\textrm{AF}\,|\,\textrm{QAF}}$ drives the investigated system towards the QAF ground state.
Obviously, the susceptibility times temperature product vanishes ($\chi T \to 0$) as temperature goes to zero ($T \to 0$) regardless of a relative strength the distortion parameter
$\delta \tilde{I}$ due to the antiferromagnetic character of both zero-field ground states AF and QAF. Moreover, it can be understood from Fig.~\ref{fig7}a that the susceptibility times temperature product increases over the whole temperature range when the distortion parameter $\delta \tilde{I}$ increases from zero up to $\tilde{I}_{\textrm{AF}\,|\,\textrm{QAF}}$. Contrary to this,
the susceptibility times temperature product increases only at relatively higher temperatures upon further strengthening of the distortion parameter from $\tilde{I}_{\textrm{AF}\,|\,\textrm{QAF}}$ to 1, while the reverse trend is generally observed at lower temperatures (see Fig.~\ref{fig7}b).

\begin{figure}[htb]
\begin{center}
\includegraphics[width=0.95\textwidth]{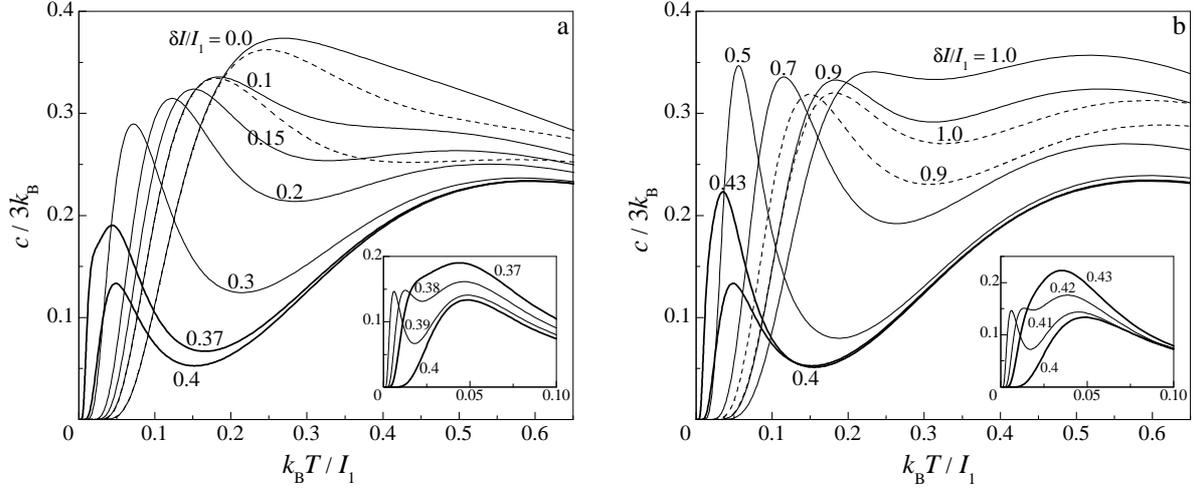}
\end{center}
\vspace{-0.5cm}
\caption{The temperature dependences of the zero-field specific heat for the antiferromagnetic Heisenberg interaction $\tilde{J}=1.5$, $\Delta=1$ ($\tilde{J}_{+}=3$)
and several values of the distortion parameter $\delta\tilde{I}$: (a) $\delta\tilde{I}\leq \delta \tilde{I}_{\textrm{AF}\,|\,\textrm{QAF}} =0.4$;
(b) $\delta\tilde{I}\geq \delta \tilde{I}_{\textrm{AF}\,|\,\textrm{QAF}}$. The dashed lines correspond to another particular case $\tilde{J}=1$, $\Delta=2$
($\tilde{J}_{+}=3$), which serve in evidence of the role of quantum fluctuations. The insets show in an enlargened scale two low-temperature peaks
emerging close to the phase boundary between the AF and QAF ground states.}
\label{fig8}
\end{figure}

Furthermore, let us explore an influence of the distortion parameter $\delta\tilde{I}$ on typical temperature dependences of the zero-field specific heat as exemplified in Fig.~\ref{fig8} on the particular example with the fixed value of the antiferromagnetic Heisenberg interaction $\tilde{J}=1.5$, $\Delta=1$ ($\tilde{J}_{+}=3$). It can be seen from Fig.~\ref{fig8}a that the zero-field specific heat displays  a standard thermal dependence with one round maximum whenever the parallelogram distortion is sufficiently small $\delta\tilde{I} \lesssim 0.15$. On the contrary, the more intriguing thermal dependence of the zero-field specific heat can be found for stronger parallelogram distortions $\delta\tilde{I} \gtrsim 0.15$. Under this condition, the zero-field specific heat exhibits at least two separate maxima originating from diverse energy scales of two different Ising couplings, whereas quantum fluctuations raised by $XY$-part of the XXZ Heisenberg coupling generally support a splitting of the emergent peaks (e.g. compare solid and dashed lines in Fig.~\ref{fig8}b). However, the most striking thermal variations of the zero-field specific heat can be detected in a close vicinity of the phase boundary $\delta\tilde{I}_{\textrm{AF}\,|\,\textrm{QAF}}$ between the AF and QAF ground states, where the additional third sharp maximum emerges at low enough temperatures (see the insets in Figs. \ref{fig8}a and b). The novel low-temperature peak can be interpreted as the Schottky-type maximum, which relates to thermal excitations from the AF ground state towards the low-lying first excited QAF state or vice versa. It actually turns out that the position of the emergent low-temperature peak moves towards lower temperatures as the distortion parameter approaches the boundary value $\delta\tilde{I}_{\textrm{AF}\,|\,\textrm{QAF}}$, whereas the height of low-temperature peak $c/3k_{\rm B} \approx 0.146$ is also in an excellent accordance with the Schottky theory for a two-level system with equal degeneracy \cite{gopal,karlova}.

\begin{figure}[h!]
\begin{center}
\includegraphics[width=0.95\textwidth]{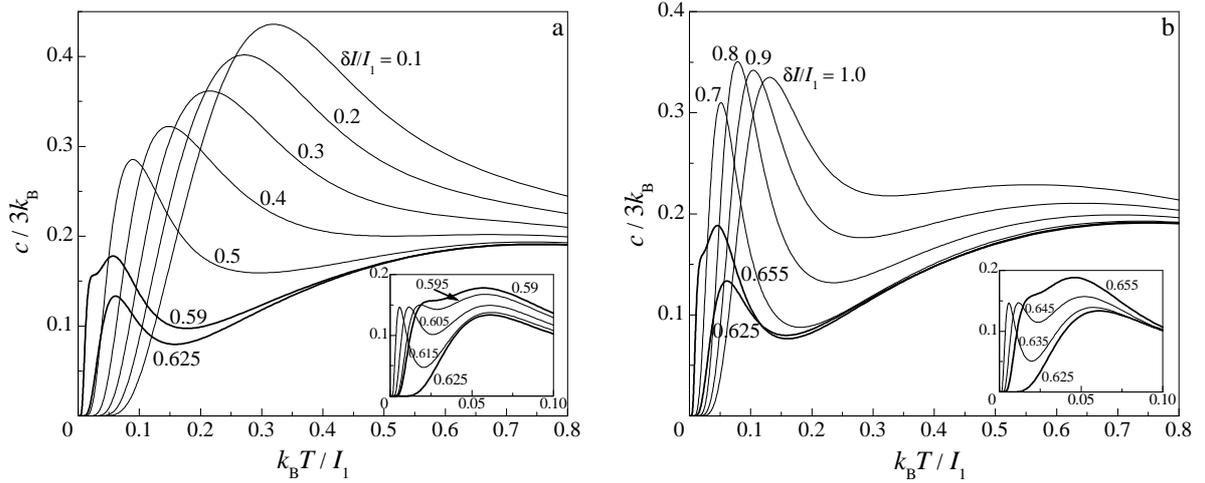}
\end{center}
\vspace{-0.5cm}
\caption{The temperature dependences of the zero-field specific heat for the ferromagnetic Heisenberg interaction $\tilde{J}=-0.5$, $\Delta=6$ ($\tilde{J}_{-}=2.5$)
and several values of the distortion parameter $\delta\tilde{I}$: (a) $\delta\tilde{I}\leq \delta \tilde{I}_{\textrm{AF}\,|\,\textrm{QAF1}} =0.625$;
(b) $\delta\tilde{I}\geq \delta \tilde{I}_{\textrm{AF}\,|\,\textrm{QAF1}}$. The insets show in an enlargened scale two low-temperature peaks
emerging close to the phase boundary between the AF and QAF1 ground states.}
\label{fig9}
\end{figure}

To gain an overall insight, the effect of parallelogram distortion on typical thermal variations of the zero-field specific heat are displayed in Fig.~\ref{fig9} on one illustrative example of the ferromagnetic Heisenberg interaction $\tilde{J}=-0.5$, $\Delta=6$ ($\tilde{J}_{-}=2.5$), which falls into the parameter region $4/(\Delta + 1)<\tilde{J}_{-} < 4$ with two different zero-field ground states AF and QAF1 depending on a relative strength of the distortion parameter (see Fig. \ref{fig4}c). Although the temperature dependences of the zero-field specific heat are quantitatively different, they undergo the same qualitative changes upon strengthening of the distortion parameter. As a matter of fact, the double-peak thermal dependences of the zero-field specific heat can be observed in Fig.~\ref{fig9} for strong enough distortion parameter $\delta\tilde{I} \gtrsim 0.3$, whereas the remarkable triple-peak thermal dependences with the relatively sharp low-temperature Schottky maximum
do emerge when the distortion parameter drives the investigated system close to the phase boundary between the AF and QAF1 ground states (i.e. $\delta \tilde{I}_{\textrm{AF}\,|\,\textrm{QAF1}}$).

At this stage, let us perform a more comprehensive analysis of the outstanding triple-peak temperature dependences of the zero-field specific heat,
which appear due to a small deviation of the distortion parameter $\delta\tilde{I}$ either from the coexistence point $\delta \tilde{I}_{\textrm{AF}\,|\,\textrm{QAF}}$
of the AF and QAF ground states (Fig.~\ref{fig8}) or the coexistence point $\delta \tilde{I}_{\textrm{AF}\,|\,\textrm{QAF1}}$ of the AF and QAF1 ground states (Fig.~\ref{fig9}).
The temperature corresponding to the maximum of the Schottky-type peak, which originates from thermal excitations between the AF and QAF (QAF1) phases,
is proportional to the respective energy difference $\tilde{\cal E}_{\textrm{AF}}-\tilde{\cal E}_{\textrm{QAF}}$ ($\tilde{\cal E}_{\textrm{AF}}-\tilde{\cal E}_{\textrm{QAF1}}$)
in a neighborhood of the coexistence point $\delta \tilde{I}_{\textrm{AF}\,|\,\textrm{QAF}}$ ($\delta \tilde{I}_{\textrm{AF}\,|\,\textrm{QAF1}}$)
\begin{eqnarray}
\left. {\tilde{\cal E}_{\textrm{AF}}-\tilde{\cal E}_{\textrm{QAF}\phantom{1}} \atop
\tilde{\cal E}_{\textrm{AF}}-\tilde{\cal E}_{\textrm{QAF1}}} \right\} \approx
\kappa_{\pm}(J,\Delta) \left(\delta \tilde{I} -
\delta \tilde{I}_{\textrm{AF}\left|{\textrm{QAF}\phantom{1} \atop \textrm{QAF1}}\right.} \right), \qquad
\kappa_{\pm}(J,\Delta) = 1 + \frac{2\delta \tilde{I}_{\textrm{AF}\left|{\textrm{QAF}\phantom{1} \atop \textrm{QAF1}}\right.}}
{\sqrt{\frac{\Delta^2}{\left(\Delta \pm 1 \right)^2} \tilde{J}_{\pm}^2 +
\left(2\delta \tilde{I}_{\textrm{AF}\left|{\textrm{QAF}\phantom{1} \atop \textrm{QAF1}}\right.}\right)^2}} .
\nonumber
\end{eqnarray}
Clearly, the interval of values $\delta \tilde{I}$ in which the specific heat exhibits three separate peaks is inversely proportional to the coefficient $\kappa_{\pm}$. Another interesting observation is that the energy difference $\tilde{\cal E}_{\textrm{AF}}-\tilde{\cal E}_{\textrm{MD}}$  for $\tilde{J}>0$ in a vicinity of the coexistence point $\delta \tilde{I}_{\textrm{AF}\,|\,\textrm{QAF}}$
is the same as the energy difference $\tilde{\cal E}_{\textrm{AF}}-\tilde{\cal E}_{\textrm{MD1}}$ for $\tilde{J}<0$
in a vicinity of the coexistence point $\delta \tilde{I}_{\textrm{AF}\,|\,\textrm{QAF1}}$
provided that
$\tilde{J}_{+}/2 + \delta \tilde{I}_{\textrm{AF}\,|\,\textrm{QAF}} =
 \tilde{J}_{-}/2 + \delta \tilde{I}_{\textrm{AF}\,|\,\textrm{QAF1}}$:
\begin{eqnarray}
\left. {\tilde{\cal E}_{\textrm{AF}}-\tilde{\cal E}_{\textrm{MD}\phantom{1}} \atop
\tilde{\cal E}_{\textrm{AF}}-\tilde{\cal E}_{\textrm{MD1}}} \right\} = \tilde{h} - 2 + \frac{1}{2} \tilde{J}_{\pm}
+ \delta \tilde{I}_{\textrm{AF}\left|{\textrm{QAF}\phantom{1} \atop \textrm{QAF1}}\right.}
+ \left(\delta \tilde{I} - \delta \tilde{I}_{\textrm{AF}\left|{\textrm{QAF}\phantom{1} \atop \textrm{QAF1}}\right.} \right).
\nonumber
\end{eqnarray}

This means that the energy spectrum composed of three states AF, QAF, and MD in a vicinity of the coexistence point $\delta \tilde{I}_{\textrm{AF}\,|\,\textrm{QAF}}$
is equivalent to the energy spectrum of three states AF, QAF1, and MD1 in a neighborhood of the other coexistence point $\delta\tilde{I}_{\textrm{AF}\,|\,\textrm{QAF1}}$.
For this reason, the low-temperature features of the heat capacity in a vicinity of both coexistence points $\delta \tilde{I}_{\textrm{AF}\,|\,\textrm{QAF}}$
and $\delta \tilde{I}_{\textrm{AF}\,|\,\textrm{QAF1}}$ are equivalent provided that
$\tilde{J}_{+}/2 + \delta \tilde{I}_{\textrm{AF}\,|\,\textrm{QAF}} =
 \tilde{J}_{-}/2 + \delta \tilde{I}_{\textrm{AF}\,|\,\textrm{QAF1}}$.
To be more specific, the particular cases of the specific heats shown in Fig.~\ref{fig8} and Fig.~\ref{fig9}
are consistent with the following values of the proportionality constants $\kappa_{+}(\tilde{J}_{+},\Delta) = 1.471$
and $\kappa_{-}(\tilde{J}_{-},\Delta) = 1.385$, respectively,
and they correspond to the following values of the coupling constants
$\tilde{J}_{+}/2 + \delta \tilde{I}_{\textrm{AF}\,|\,\textrm{QAF}} = 1.9$ and
$\tilde{J}_{-}/2 + \delta \tilde{I}_{\textrm{AF}\,|\,\textrm{QAF1}}= 1.875$.
Owing to these facts, the specific heat displayed in Fig.~\ref{fig8} for the antiferromagnetic Heisenberg coupling $\tilde{J}=1.5$, $\Delta=1$ ($\tilde{J}_{+}=3$) has three separate peaks in a slightly smaller interval of the distortion parameter $\delta \tilde{I}$ than the specific heat shown in Fig.~\ref{fig9} for the ferromagnetic Heisenberg interaction $\tilde{J}=-0.5$, $\Delta=6$ ($\tilde{J}_{-}=2.5$).

\begin{figure}[htb]
\begin{center}
\includegraphics[width=0.95\textwidth]{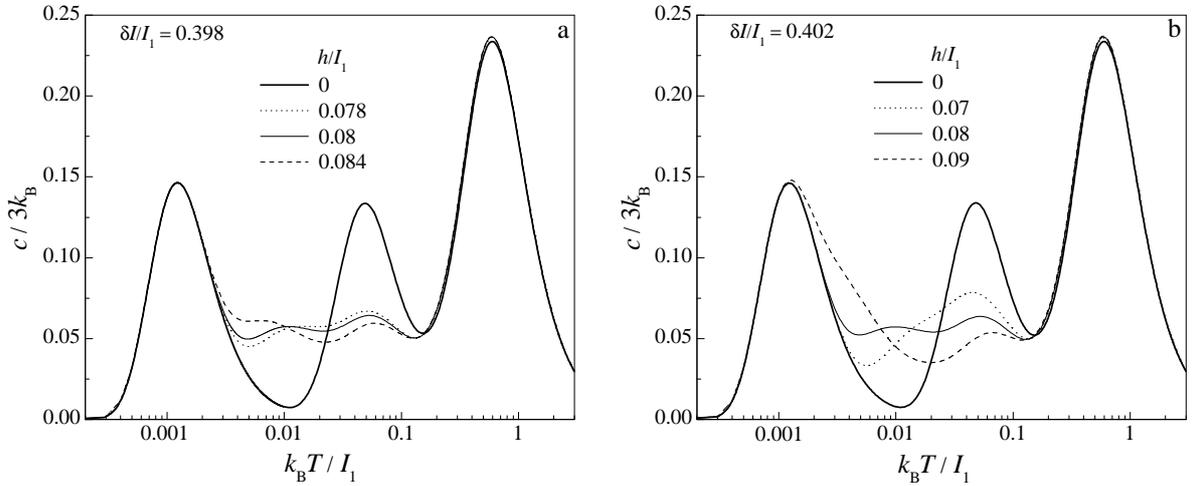}
\end{center}
\vspace{-0.5cm}
\caption{The semi-logarithmic plot of temperature dependences of the specific heat for the antiferromagnetic Heisenberg interaction $\tilde{J}=1.5$, $\Delta=1$ ($\tilde{J}_{+}=3$),
several values of the magnetic field $\tilde{h}$ and two different values of the distortion parameter $\delta\tilde{I}$ selected close to the coexistence
point $\delta \tilde{I}_{\textrm{AF}\,|\,\textrm{QAF}}=0.4$: (a) $\delta\tilde{I}=0.398$; (b) $\delta\tilde{I}=0.402$.}
\label{fig10}
\end{figure}
Last but not least, let us examine temperature variations of the specific heat in a presence of non-zero external magnetic field. The most interesting thermal variations of the zero-field specific heat has been formerly found in a close neighborhood of the coexistence point $\delta \tilde{I}_{\textrm{AF}\,|\,\textrm{QAF}}$ between the AF and QAF ground states and hence, our primary attention will be therefore paid to this parameter region. It can be seen from Fig.~\ref{fig10} that a suitable choice of the distortion parameter and small magnetic field gives rise to a remarkable temperature dependence of the specific heat with up to four separate maxima -- the main and three additional. The three out of four peaks of the specific heat already emerge at zero magnetic field.  While the round maximum at the highest temperature involves thermal excitations of diverse physical origin, the preferred thermal excitations in between the AF and QAF states can be entirely connected with the Schottky-type maximum emerging at the lowest temperature.
The third subtle maximum, which can be observed in the zero-field specific heat at moderate temperatures, originates from other preferential thermal excitations from two lowest-energy AF and QAF states towards low-lying excited states arising out from the FRU state. It is quite obvious from Fig.~\ref{fig10} that the height of the moderate maximum is rapidly suppressed by a relatively small magnetic field, whereas the moderate maximum simultaneously splits into two less marked maxima. The outstanding splitting of the moderate peak at a relatively small magnetic field can be attributed to the Zeeman's splitting of available spin states of the nodal Ising spins within the highly degenerate FRU states.
Finally, it is worthwhile to remember that the similar temperature dependence of the specific heat with four distinct peaks has been already reported for the undistorted mixed spin-(1,1/2) Ising--Heisenberg diamond chain ($\delta \tilde{I}=0$), but it came into being due to the combined effect of the uniaxial single-ion anisotropy and the magnetic field \cite{jmmm15}.

\section{Conclusion}
\label{conclusion}

In the present article we have rigorously examined the ground state and thermodynamics of the mixed spin-(1,1/2) Ising--Heisenberg distorted diamond chain within the transfer-matrix method. In particular, our attention was focused on how the parallelogram distortion affects the magnetization process, susceptibility and specific heat of the mixed spin-(1,1/2) Ising--Heisenberg
distorted diamond chain with the antiferromagnetic Ising interactions and either the antiferromagnetic or ferromagnetic XXZ Heisenberg interaction. Under this circumstances, the magnetic properties of the mixed spin-(1,1/2) Ising--Heisenberg distorted diamond chain are substantially influenced by a geometric spin frustration.

The ground-state phase diagram of the mixed spin-(1,1/2) Ising--Heisenberg distorted diamond chain with the antiferromagnetic (ferromagnetic) Heisenberg interaction totally consists of
four different ground states: the saturated paramagnetic state SPA, the classical antiferromagnetic state AF, the monomer-dimer state MD (MD1) and the quantum antiferromagnetic state QAF (QAF1).
The quantum ground states MD and QAF emerging for the antiferromagnetic Heisenberg interaction differ from the analogous quantum ground states MD1 and QAF1 emerging for the ferromagnetic Heisenberg interaction just by antisymmetric and symmetric quantum superposition of two antiferromagnetic states of the Heisenberg spin pairs, respectively.
The ground-state phase diagram in the distortion parameter -- magnetic field $(\delta\tilde{I}, \tilde{h})$ plane can have two different topologies depending on the antiferromagnetic Heisenberg coupling $\tilde{J}_{+}=\tilde{J}\left(\Delta + 1 \right)$ and four different topologies depending on the ferromagnetic Heisenberg coupling $\tilde{J}_{-}=|\tilde{J}|\left(\Delta - 1 \right)$.

It has been demonstrated that the magnetization curve of the mixed spin-(1,1/2) Ising--Heisenberg distorted diamond chain may involve at most two different intermediate plateaus at zero and one-half of the saturation magnetization. From this perspective, the distortion parameter does not lead to a creation of novel magnetization plateaus in comparison with the undistorted case \cite{jmmm15}. On the other hand, the distortion parameter is responsible for a rich variety of temperature dependences of the specific heat, which may display one, two or three anomalous low-temperature peaks in addition to the round maximum observable at higher temperatures. The physical origin of all observed low-temperature peaks of the specific-heat has been clarified on the grounds of preferred thermal excitations. It is worthwhile to remark that the investigated spin system reduces to the mixed spin-(1,1/2) Ising--Heisenberg doubly decorated chain in the particular case $I_2=0$ ($\delta \tilde{I}=1$) and the symmetric mixed spin-(1,1/2) Ising--Heisenberg diamond chain in the other particular case $I_1=I_2$ ($\delta \tilde{I}=0$) \cite{jmmm15}.

\end{document}